\documentclass[preprint,12pt,longnamesfirst]{aastex}
\usepackage{epsfig}
\def\um{$\mu$m}
\begin{document}
\title{{\it Spitzer} 24 \um\ Images of Planetary Nebulae}
%
%
\author{You-Hua Chu\altaffilmark{1}, Robert A.\ Gruendl\altaffilmark{1},
Martin A.\ Guerrero\altaffilmark{2}, Kate Y.~L.\ Su\altaffilmark{3}, 
Jana Bilikova\altaffilmark{1}, Martin Cohen\altaffilmark{4},
Quentin A.\ Parker\altaffilmark{5},  Kevin Volk\altaffilmark{6},
Adeline Caulet\altaffilmark{1},
Wen-Ping Chen\altaffilmark{7}, Joseph L.\ Hora\altaffilmark{8},  
Thomas Rauch\altaffilmark{9}}
\altaffiltext{1}{\itshape Department of Astronomy, University of Illinois
at Urbana-Champaign, 1002 West Green Street, Urbana, IL 61801;
chu@astro.uiuc.edu} 
\altaffiltext{2}{\itshape Instituto de Astrof\'{\i}sica de Andaluc\'{\i}a,
CSIC. c/ Camino Bajo de Hu\'etor 50, E-18008 Granada, Spain}
\altaffiltext{3}{\itshape Stewart Observatory, University of Arizona,
Tucson, AZ 85721}  
\altaffiltext{4}{\itshape Radio Astronomy Laboratory, University of 
California, Berkeley, CA 94720} 
\altaffiltext{5}{\itshape Department of Physics, Macquarie University, 
Sydney, NSW 2109, Australia} 
\altaffiltext{6}{\itshape Space Telescope Science Institute,
3700 San Martin Drive, Baltimore, MD 21218} 
\altaffiltext{7}{\itshape Institute of Astronomy, National Central 
University, Chung-Li, Taiwan} 
\altaffiltext{8}{\itshape Center for Astrophysics, 60 Garden St., MS 65,
Cambridge, MA 02138} 
\altaffiltext{9}{\itshape Institut f\"ur Astronomie und Astrophysik
 T\"ubingen (IAAT), Abteilung Astronomie, Sand 1, D-72076 T\"ubingen,
Germany} 
%
%

%

%
%
\begin{abstract}
{\it Spitzer} MIPS 24 \um\ images were obtained for 36 Galactic 
planetary nebulae (PNe) whose central stars are hot white dwarfs
(WDs) or pre-WDs with effective temperatures of $\sim$100,000 K 
or higher.  Diffuse 24 \um\ 
emission is detected in 28 of these PNe.  The eight non-detections 
are angularly large PNe with very low H$\alpha$ surface brightnesses.
We find three types of correspondence between the 24 \um\ emission
and H$\alpha$ line emission of these PNe: six show 24 $\mu$m emission
more extended than H$\alpha$ emission, nine have a similar extent at
24 \um\ and H$\alpha$, and 13 show diffuse 24 \um\ emission near the 
center of the H$\alpha$ shell.
The sizes and surface brightnesses of these three groups of PNe and
the non-detections suggest an evolutionary sequence, with the
youngest ones being brightest and the most evolved ones undetected.
The 24 \um\ band emission from these PNe is attributed to 
[\ion{O}{4}] 25.9 \um\ and [\ion{Ne}{5}] 24.3 \um\ line emission
and dust continuum emission, but the relative contributions of these
three components depend on the temperature of the central star and
the distribution of gas and dust in the nebula.
\end{abstract}
\subjectheadings{circumstellar matter -- infrared: stars --
 planetary nebulae: general -- white dwarfs}
\maketitle
%
%
%
\section{Introduction}  \label{sec:intro}

Low- and intermediate-mass stars expel their envelopes to form
planetary nebulae (PNe), while their exposed cores evolve into 
white dwarfs (WDs).  The morphology of a PN can be very complex.  
First of all, a PN may show a multiple-shell morphology due to 
interactions between fast and slow winds \citep{Kwok83}
and between the winds and the ambient interstellar medium 
\citep{DS98,Vetal03}.
A bipolar structure may be developed in a PN owing to a fast 
stellar rotation or close binary interaction \citep{SL89}.  
Moreover, the effective temperature of the central star of a PN 
(CSPN) increases as the nebula expands, and the varying 
stellar UV flux results in varying ionization and excitation 
structure in the PN \citep{Schon83}.  Therefore, images of a 
PN taken at different wavelengths may exhibit very different
morphologies, and comparisons among these images help us 
understand the PN's physical conditions and structure.

Infrared (IR) imaging of PNe from the ground is easy at 
wavelengths $\lesssim$2 \um, but is hampered by the bright sky
background at longer wavelengths.  While high-resolution images
at wavelengths up to $\sim$20 \um\ can be obtained from the 
ground for the bright, young, compact PNe 
\citep[e.g.,][]{Vetal07}, it is difficult to observe extended
PNe with lower surface brightnesses.  The {\it Spitzer 
Space Telescope} \citep{Werner04} makes it possible for the 
first time to obtain high-sensitivity images of a large number 
of older and more extended PNe at near- and mid-IR wavelengths.  
Its InfraRed Array Camera \citep[IRAC;][]{Fazio04} takes images 
at 3.6, 4.5, 5.8, and 8.0 \um, while the Multiband Imaging 
Photometer for {\it Spitzer} \citep[MIPS;][]{Rieke04} takes
images at 24, 70, and 160 \um.  
IRAC images are available for over 60 PNe and 
they have detected emission from ionized gas, warm dust, 
polycyclic aromatic hydrocarbons (PAHs), and H$_2$ molecular 
gas \citep[e.g.,][]{Hora04,Hetal08,Hora09}.
MIPS images, before our work, were available for only about a 
dozen mature PNe, excluding proto-PNe that were usually unresolved
or saturated in MIPS 24 \um\ images.
Analyses of MIPS observations of NGC\,650 \citep{Ueta06}, 
NGC 2346 \citep{Su04}, and NGC 7293 \citep{Su07} suggest that 
diffuse emission in the 24 \um\ band is dominated by the 
[\ion{O}{4}] 25.9 \um\ line, and the 70 and 160 \um\ bands by 
dust continuum emission.

We have conducted a {\it Spitzer} MIPS 24 \um\ survey of 72 hot
WDs or pre-WDs, selected from \citet{MS99} and \citet{Nap99}, to 
search for dust disks at radii of a few tens of AU (PID 40953).  
Thirty-six of these hot stars surveyed are still surrounded by 
visible PNe; therefore, our MIPS survey of hot WDs has provided 
a three-fold increase of 24 \um\ images of spatially resolved PNe 
in the entire {\it Spitzer} archive.
Diffuse emission in the 24 \um\ band is detected from 28 PNe, among
which $\sim$50\% show shell morphologies and $\sim$35\% show emission
only in the central region near the CSPN.
In this paper, we describe the observations in Section 2, present
the results in Section 3, and discuss the origins of the 24 \um\ emission
in Section 4 and implications of this emission in Section 5.

\section{Observations}  \label{sec:obs}

The MIPS
onboard the {\it Spitzer Space Telescope} was used to image the 36 PNe
in the 24 \um\ band as part of Program 40953.  Each object was imaged in 
the photometry mode using the small offset scale with 10~s exposure time
for three cycles.  The Basic Calibrated Data processed by the {\it Spitzer} 
Science Center's pipeline software were used to produce mosaics of the PNe.
Prior to building each mosaic, bad pixels and latent images of bright
point sources in each frame were flagged, and background brightness 
offsets between individual frames were determined and corrected using 
the method outlined by \citet{RG95}.  The MOPEX software package was
used to combine data and produce the mosaics.
The field-of-view of each mosaic covers $\sim$8.0$'$$\times$7.5$'$,
of which the central $\sim$3.5$'$$\times$3.0$'$ portion has a total
exposure time of $\sim$420~s and the outer regions have exposure times
1/2 or 1/4 that of the center.  We note that latent images of the 
brighter PNe sometimes persist in the outer portions of the final 
mosaics.  The point-spread function of MIPS at 24 \um\ has a FWHM
of $\sim$6$''$.

\section{Results}  \label{sec:results}

Among the 36 PNe that have been imaged in the MIPS 24 \um\
band, 28 are detected and exhibit a variety of morphologies.  
To understand the origin of this emission,
we compare each PN's 24 \um\ morphology to its optical morphology.
H$\alpha$ images, showing all dense ionized gas, are best for
such comparison; however, not all PNe have H$\alpha$ 
images\footnote{H$\alpha$ images of MeWe 1-3, HeDr 1,
Lo 4, and BlDz 1 are taken from AAO/UKST H$\alpha$ Survey
\citep{Petal05}; 
those of MWP 1, NGC 246, and Sh 2-188 were
obtained using the Lulin Observatory 1 m telescope; that
of NGC 1360 was obtained with the Mt.\ Laguna 1 m telescope;
those of Abell 21, Abell 39, Abell 43, EGB 1, IC 1295, 
Jn 1, and JnEr 1 are taken from The IAC Catalog of Northern 
Galactic Planetary Nebulae \citep{Man96}; those 
of IC 289, NGC 2438, NGC 2610, NGC 3587, and NGC 6852 
\citep[published by][]{Hetal97,Corradi00,Corradi03} have 
been downloaded from the Image Database of Planetary Nebulae 
Haloes Web page maintained by Romano L.\ M.\ Corradi.}
available, and we have to resort to
the Digitized Sky Survey 2 red (DSS2r) images.
For PNe with hot CSPNs, the nebular excitation and ionization
is high and the strongest emission line in the DSS red passband
is the H$\alpha$ line; thus DSS2r images are excellent substitutes
for H$\alpha$ images.
In one exception, HaTr 7, the DSS 2 blue (DSS2b) image is used
because it shows the nebula better than the DSS2r image \citep{HT85}.
For the rest of the paper, we will loosely use these DSS2r and 
DSS2b images as ``H$\alpha$'' images, assuming that they are similar.

Table 1 lists all 36 PNe observed:
Column 1, PN name;
Column 2, Galactic coordinates;
Column 3, CSPN name;
Column 4, the stellar effective temperature ($T_{\rm eff}$); 
Column 5, the angular size (diameter or major $\times$ minor axes) 
          of the bright main nebula measured from the H$\alpha$ images;
Column 6, distance; 
Column 7, the linear size of the main nebula in H$\alpha$; 
Column 8, the \ion{He}{2} $\lambda$4686/H$\beta$ flux ratio; and
Column 9, the peak surface brightness at 24 $\mu$m.
The eight PNe that were not detected in the 24 $\mu$m images are
listed at the end of Table 1.

To visualize the spatial distribution of the 24 \um\ and H$\alpha$ 
emission, we have extracted surface brightness profiles along cuts 
through the nebular center (i.e., the CSPN) at position angles (PAs)
selected to avoid contaminating stars and to sample the well-defined
shell rims of the main nebula.
The width of each cut, typically a few arcsec, has been adjusted to
achieve a reasonable signal-to-noise ratio (S/N) in the resulting profile.
For each PN, similar widths are used to extract the 24 \um\ and 
H$\alpha$ surface brightness profiles, and the profiles are
individualy zeroed at the background level and normalized to 
the peak of nebular emission.
Note that the H$\alpha$ surface brightness profiles are not
normalized in the cases of Abell 61, HaTr 7, and Sh 2-216, because
of stellar contamination in the first object and extremely low
S/N in the images of the latter two PNe.

Using the images and surface brightness profiles, we find three 
kinds of correspondence between the 24 \um\ and H$\alpha$ 
morphologies of the 28 PNe:
(1) the 24 \um\ emission is more extended than the H$\alpha$ emission,
(2) the 24 \um\ emission has a similar spatial extent as the H$\alpha$
    emission, and
(3) the 24 \um\ emission peaks near the CSPN within the H$\alpha$ shell.
We have used these different correspondences between 24 \um\
and H$\alpha$ emission to divide the PNe into three groups
accordingly, and arranged them in Table 1 and Figures 1--3 in the
order of these groups to facilitate easy comparisons among objects.
Some PNe have properties intermediate between Groups 1 and 2, and
they are assigned to groups based on larger likelihoods.
More detailed descriptions of these groups are given below.

\subsection{Group 1}

PNe in Group 1 (Figure 1) appear more extended in MIPS 24 \um\ images 
than in H$\alpha$ images.  Their images also show sharper outer
rims in H$\alpha$ than at 24 \um, but the differences in sharpness 
are in part caused by different angular resolutions, $\sim$6$''$
for 24 \um\ and $\le$2.5$''$ for H$\alpha$.
The spatial extent of emission is better illustrated by the
surface brightness profiles. 
Outside the bright shell rim, the 24 \um\ surface brightness falls 
off more slowly than the H$\alpha$ surface brighness, resulting in
a vertical offset between the two profiles in Figure 1.
The 24 \um\ emission is also 10--20$''$ more extended than the 
H$\alpha$ emission.
These differences cannot be attributed to the poorer instrumental 
resolution of the 24 $\mu$m images.
To further illustrate this point, we have convolved the DSS2r image of
Abell 15 with a Gaussian of 6$''$ FWHM, the resolution of the MIPS 
24 \um\ PSF, and plotted the convolved surface brightness profile in 
a dotted curve in Figure 1.
It is clear that the convolved ``H$\alpha$'' surface brightness profile 
is still not as extended as the 24 \um\ surface brightness profile. 

PNe in Group 1 all show similar morphologies at 24 \um: a bright inner
shell surrounded by a faint outer envelope; however, the outer envelopes
have obvious optical counterparts only in IC 289 and NGC 2610.  In the 
case of NGC 6852, the 24 \um\ image shows a smooth spherical outer envelope, 
but its H$\alpha$ and [\ion{O}{3}] images show faint bipolar outer 
features along PAs of 140$^\circ$ and 320$^\circ$ \citep{Man96}.

\subsection{Group 2 }

PNe in Group 2 (Figure 2) have 24 \um\ and H$\alpha$ surface brightness 
profiles falling off in a similar fashion at the outer rim.
Some of these PNe have a well-defined shell morphology, e.g.,
Abell 39, Abell 43, and NGC 246, but some have irregular shapes
or a centrally filled morphology, e.g., HeDr 1 and MWP 1.
Overall the morphologies are similar at 24 \um\ and H$\alpha$, although
the details may differ.
For example, Jn 1 shows clear limb-brightening in H$\alpha$, but
appears more filled in at 24 \um.
NGC 246 shows enhanced 24 \um\ emission at 40$''$--45$''$ NW and SE of
the CSPN, and its IRAC images also show displaced enhancements along
the same PAs \citep{Hora04}, but no H$\alpha$ counterparts are seen.

\subsection{Group 3}

PNe in Group 3 (Figure 3) show diffuse 24 \um\ nebular emission 
only in the central region near the CSPN, totally different from
the distribution of H$\alpha$ emission.
In the case of NGC\,1360, the H$\alpha$ and 24 \um\ morphologies
are not strikingly different, but the 24 \um\ surface brightness
profile is much more centrally peaked than the H$\alpha$ profile.
A variety of H$\alpha$ morphologies are observed in this group 
of PNe: thick shells in JnEr 1 and NGC 2438; centrally
filled shells in Abell 61, BlDz 1, HaTr 7, IC 1295, K 1-22, 
NGC 1360, NGC 3587, and Sh 2-216; semi-circular arcs in Abell 21 
and Sh 2-188; and irregular morphology in EGB 1.
It is interesting to note that IC 1295, K 1-22, and NGC 3587
show 24 \um\ emission only near the center of the H$\alpha$ shell,
but the 24 $\mu$m morphology is strikingly similar to a scaled-down
H$\alpha$ morphology.  In all three cases, the 24 \um\ image is 
bisected by a band of lower surface brightness along the major axis 
of the H$\alpha$ shell.

\subsection{Nondetections}

Eight PNe do not show detectable nebular emission in the 
24 \um\ band.  Seven of these PNe are angularly larger than
the areas that were mapped by the MIPS 24 \um\ observations 
(see Column 5 of Table 1), and their surface brightnesses are 
very low near the CSPNs, where the MIPS observations were centered.
There might be some diffuse 24 \um\ emission in the PN
JavdSt 1 \citep{Jacoby95}, but it is close to the noise level.  
BE UMa is a cataclysmic variable with a $\sim$3$'$ nebular shell
\citep{Liebert95}, which is completely covered by the MIPS 24 \um\ 
mosaic, but no diffuse 24 \um\ emission is detected.

\section{Origins of the MIPS 24 \um\  Emission}  \label{sec:discussion}

What are the origins of the diffuse emission detected in the 
{\it Spitzer} MIPS 24 \um\ band?
The [\ion{Ne}{5}] 24.3 \um\ and [\ion{O}{4}] 25.9 \um\ nebular lines
and dust continuum are the three main candidates contributing to 
the MIPS 24 \um\ fluxes, as indicated by {\it Infrared Space Observatory}
({\it  ISO}) SWS spectra of the 
young PN NGC 7027 and the bipolar PN NGC 2346.
In NGC 7027, [\ion{O}{4}] and [\ion{Ne}{5}] lines and dust 
continuum are all significant, although the [\ion{Ne}{5}]
line is only 1/2 as bright as the [\ion{O}{4}] line \citep{BS01}.
In NGC 2346, the [\ion{O}{4}] line contributes 27\%, and dust 
continuum contributes the rest of the emission in the MIPS 24 
\um\ band \citep{Su04}.
The excitation potentials of \ion{Ne}{5} and \ion{O}{4} are
97.1 and 54.9 eV, respectively.
It is not surprising that NGC 7027 and NGC 2346 have such different
[\ion{Ne}{5}]/[\ion{O}{4}] ratios, as the former CSPN has an 
effective temperature of $\sim$160,000 K \citep{Betal96} and 
the latter CSPN has a Zanstra temperature of only 
$\sim$100,000 K \citep{Men78}.
It is conceivable that the [\ion{Ne}{5}]/[\ion{O}{4}] ratio 
increases with stellar temperature, and for CSPNs with effective 
temperatures $\le$150,000 K the [\ion{O}{4}] 25.9 \um\ line is 
by far the strongest emission line in the MIPS 24 $\mu$m band.

The ubiquitous presence of [\ion{O}{4}] line emission in
our PNe is indicated by the \ion{He}{2} $\lambda$4686 emission 
that is detected in all available spectroscopic observations of 
these PNe \citep[see Column 8 of Table 1; compiled by][]{Tylenda94}.
The \ion{He}{2} line is emitted through recombinations of 
\ion{He}{3}, of which the excitation potential is 54.4 eV.
This excitation potential is close to that of \ion{O}{4}, 
thus wherever the \ion{He}{2} emission is present there ought 
to be [\ion{O}{4}] emission, as long as the density is not
excessively higher than the critical density of [\ion{O}{4}],
$\sim9\times10^3$ cm$^{-3}$.

Note that the \ion{He}{2}/H$\beta$ ratios in Column 8 of Table 1
appear to be higher for PNe in Groups 1 and 2 than for those in Group 3,
but this is an artificial effect of how the observations
were conducted.
Optical spectrophotometric observations of PNe are usually 
made in regions of high surface brightness.  For PNe in
Groups 1 and 2, the optically bright regions are also
bright in the MIPS 24 \um\ band; however, for PNe in Group 3, 
the bright 24 \um\ emission peaks near the center of a nebula, 
offset from the brightest optical emission in the shell where
spectrophotometric observations are made. 
The \ion{He}{2}/H$\beta$ ratios at the central diffuse 
24 $\mu$m emission regions should be much higher than those 
in the nebular shells for Group 3 PNe.
The radial variations of the \ion{He}{2}/H$\beta$ ratio,
as well as the [\ion{Ne}{5}]/[\ion{O}{4}] ratio, reflect the 
well-known excitation and ionization stratifications in PNe.

The relative importance of nebular line emission and dust 
continuum in the MIPS 24 \um\ band depends on the ionization
structure and dust content of a PN.
It has been illustrated by {\it Spitzer} IRS observations of 
PNe in the Large and Small Magellanic Clouds that bipolar PNe
are generally dustier than round/elliptical PNe and that the   
largest PNe exhibit the least dust continuum emission \citep{Stan07}.
Group 1 PNe have 24 \um\ emission more extended than the H$\alpha$ 
emission.  As [\ion{O}{4}] tracks \ion{He}{2} and the
\ion{He}{2}/H$\alpha$ ratio decreases with radial distance from
the CSPN, due to ionization and excitation stratification, 
the extended emission in the MIPS 24 \um\ band must be dominated by
dust continuum.
For the main shells of PNe in Groups 1 and 2, it is likely
that both the nebular [\ion{O}{4}] line emission and dust continuum 
contribute to the MIPS 24 \um\ flux, but the exact ratios
cannot be determined from the current data; spectroscopic 
observations are needed.

To determine the ratio of line emission to dust continuum for
Group 3 PNe, we take a detailed look at {\it Spitzer} observations 
of diffuse emission from the Helix Nebula \citep{Su07}.
The IRAC images and MIPS 70 and 160 \um\ images of the Helix 
Nebula show the prominent helical ring structure with a hollow 
center, while the MIPS 24 \um\ image shows bright diffuse 
emission in the central region but only faint emission in the 
helical ring. 
{\it Spitzer} IRS observations were made at the central star
and its immediate surroundings without additional observations
of an external background region.
These observations were adequate for deriving a spectrum of the
central star but not a spectrum of the nebular emission because
the zodiacal background was bright toward the Helix Nebula.
MIPS 24 \um\ observations indicated that the nebular surface 
brightness was only about 1/7 that of the zodiacal emission.
Indeed, the spectrum of a diffuse emission region near the Helix
central star, shown in Figure 4, is dominated by the zodiacal 
dust continuum emission, and it is not clear whether a small
fraction of the dust continuum originates in the Helix Nebula.
Nevertheless, the nebular spectrum in Figure 4 shows strong
[\ion{O}{4}] 25.9 \um\ line emission; the [\ion{Ne}{5}] 24.3 
\um\ line is much weaker than that of the region encompassing
the central star (see Figure 2 of Su et al.\ 2007).

Recently, we obtained {\it Spitzer} IRS observations of 
the CSPN of Sh 2-188 in the mapping mode (PID 50629).
While these observations will be presented in another paper,
we have used these data to extract a spectrum of the
diffuse emission near the central cavity.
Figure 5 shows the slit positions marked on a MIPS 24 $\mu$m
image of Sh 2-188.
The spectra extracted from the nebular aperture ``Neb'' and from
the background aperture ``Bkg'' show significant amounts of dust
continuum emission and PAH emission.
The origins of this dust continuum and PAH emission are not
clear and could be nebular, interstellar, or even zodiacal.
The background-subtracted nebular spectrum shows essentially 
all fine-structure line emission, with the [\ion{O}{4}] line
dominating in the MIPS 24 $\mu$m band.
Therefore, at least in the case of Group 3 PN Sh 2-188 the 
central diffuse emission in the MIPS 24 $\mu$m band is 
by far dominated by the [\ion{O}{4}] line emission.
It is likely that the central diffuse 24 $\mu$m emission
from all Group 3 PNe is dominated by the [\ion{O}{4}] line.
The contribution of [\ion{Ne}{5}] line is smaller and
depends on the effective temperature of the CSPN and 
the radial distance to the CSPN.

Finally, we note a complication in NGC 2438.  While NGC 2438
is placed in Group 3 with centrally peaked 24 \um\ emission,
diffuse 24 \um\ emission at lower levels is detected from
its nebular shell as well as its halos.  As shown in 
Figure 6, an archival IRAC 8 \um\ image of NGC 2438 has detected 
two halos around its main nebular shell, similar to the 
H$\alpha$+[\ion{N}{2}] image reported by \citet{Corradi03};
our MIPS 24 \um\ image of NGC 2438 has detected the main
nebular shell and the inner halo clearly, and perhaps some
faint emission from the outer halo as well.
The IRAC 8 \um\ band contains strong PAH emission features
at 7.7 and 8.6 \um, forbidden lines of [\ion{Ar}{2}], 
[\ion{Ar}{3}], [\ion{Ar}{5}], and [\ion{Ne}{6}],
and weak \ion{H}{1} recombination lines, such as Pf$\alpha$.
Since these halos are detected in all IRAC bands, including
the 4.5 \um\ band that does not include PAH features,
it is likely that line emission dominates the near-IR IRAC
bands.
The 24 \um\ image of NGC 2438 resembles its [\ion{O}{3}]
$\lambda$5007 image \citep{Corradi03}, suggesting that 
[\ion{O}{4}] 25.9 \um\ line emission may make significant 
contributions to the MIPS 24 \um\ flux.  
The relative importance of nebular line and dust continuum
emission at 24 \um\ is unknown.
NGC 2438 is projected near the open cluster M46, but their
different radial velocities suggest that they are unrelated
\citep{Ketal08}.

\section{Implications for PN Evolution}  \label{sec:implication}

We have defined three groups of PNe according to the spatial 
extent and surface brightness profile of the 24 \um\ emission 
relative to those of the H$\alpha$ emission.
To compare physical properties among these three groups of PNe,
we have used the information in Table 1 and plotted the distribution 
of stellar $T_{\rm eff}$ and nebular sizes in Figure 7.
A rough trend of size increase is seen from Group 1 through
Groups 2 and 3 to the nondetections, suggesting an evolutionary
sequence.
The stellar $T_{\rm eff}$, on the other hand, does not show any
obvious trend.  This is not surprising because stellar evolutionary
tracks bend at the high-temperature extremes and different initial 
stellar masses lead to different extreme stellar temperatures;
thus stellar temperature distributions are not sensitive indicators
of evolution.

We have measured the peak 24 \um\ surface 
brightness in the shell rim of each PN in Groups 1 and 2 and 
in the central diffuse-emission region of each PN in Group 3.  
These surface brightness measurements are listed in Column 9 of Table 1.  
Two Group 2 PNe, Lo 4 and NGC 246, have peculiar bright spots 
within their nebular shells; the surface brightnesses of these 
bright spots are about twice as high as the peak surface brightness
in the shell rim. 
For the PNe that have distances available, we plot their 24 \um\ 
surface brightness versus their linear optical sizes in Figure 8.
It is evident that the 24 \um\ surface brightness decreases
with nebular size.
This correlation is expected from nebular evolution, as
nebular expansion leads to lower density and lower surface
brightness.
Column 9 of Table 1 shows that Group 1 PNe on average have the
highest surface brightnesses, and Group 3 PNe have the lowest 
surface brightnesses.  The PNe that do not show detectable 
diffuse 24 \um\ emission are still fainter than Group 3 PNe.
The trend that the 24 \um\ surface brightness decreases from 
Group 1 through Groups 2--3 to the non-detections provides 
further support of an evolutionary sequence among these PNe.

\section{Summary}

We have obtained \emph{Spitzer} MIPS 24 \um\ observations of 36 
Galactic PNe, and detected diffuse emission in 28 of them.
Comparing the 24 \um\ surface brightness profiles with those 
in the H$\alpha$ line, we find the 24 $\mu$m emission can be
more extended, similar, or present only in the shell center,
and divided these PNe into Groups 1--3, respectively.
The optical shell sizes and 24 $\mu$m surface brightnesses 
of the PNe observed suggest that Groups 1--3 and non-detections 
represent an age or evolutionary sequence, with Group 1 being 
the youngest.

The diffuse emission in the MIPS 24 $\mu$m band is attributed 
to [\ion{O}{4}] 25.9 \um\ and [\ion{Ne}{5}] 24.3 \um\ lines 
and dust continuum emission, but the relative proportions of 
these three components depend on the nebular excitation 
and dust content. 
The extended 24 \um\ emission in Group 1 PNe is most likely
dominated by dust continuum, as high-ionization fine-structure
line emission is expected to decline radially outward.
Spectroscopic observations are needed to determine the
ratio of nebular line emission to dust continuum for the
main bodies of PNe in Groups 1 and 2.
Using a {\it Spitzer} IRS spectrum of the central diffuse
emission region of PN Sh 2-188, we illustrate that the
[\ion{O}{4}] line is by far the dominant contributor
to the MIPS 24 \um\ emission.  
It is likely that [\ion{O}{4}] line emission, and some 
[\ion{Ne}{5}] line emission for PNe ionized by the hottest 
central stars, dominates the central diffuse  24 \um\ 
emission among all Group 3 PNe.

\acknowledgments
This research was supported by NASA grants JPL 1319342
and JPL 1343946. 
MAG acknowledges support by grants AYA2005-01495 of the 
Spanish Ministerio de Educaci\'on y Ciencia (MEC) and 
AYA2008-01934 of the Spanish Ministerio de Ciencia e 
Innovaci\'on (MICINN).
We thank the anonymous referee for helpful comments.

\clearpage
\begin{figure}
\epsscale{0.8}
\plotone{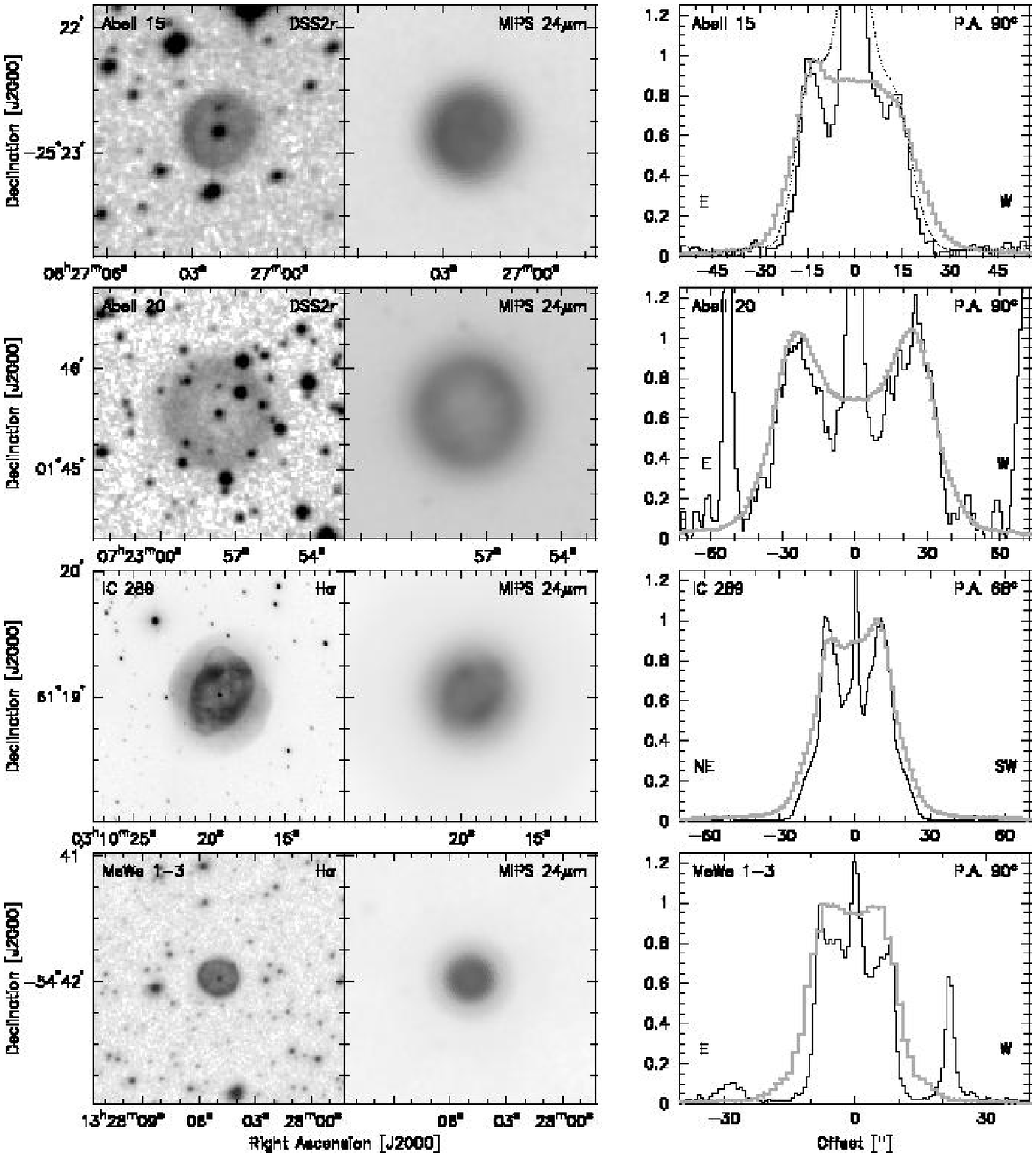}
\end{figure}
\begin{figure}
\figurenum{1}
\plotone{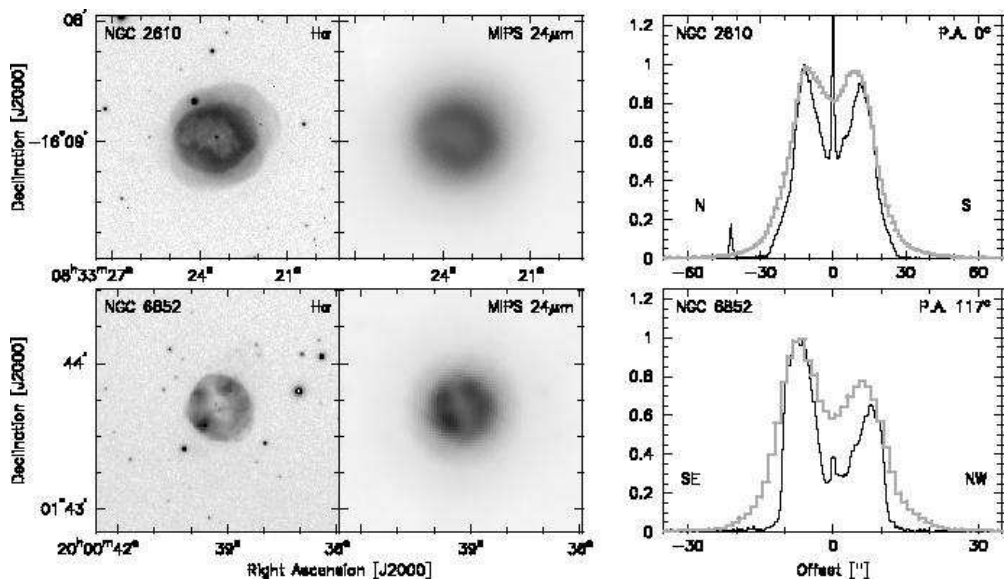}
\label{fig1}
\caption{Images and surface brightness profiles of Group 1 PNe.
The nebula names and passbands are marked on the upper left and
right corners, respectively.  The surface brightness profiles are
extracted through the nebular center along the position angle PA
labeled at the upper right corner.  The H$\alpha$ surface brightness
profile is plotted in black and 24 $\mu$m in grey.  Note that
the ``H$\alpha$'' surface brightness profiles of Abell 15 and 
Abell 20 are extracted from DSS2r images.
The ``H$\alpha$'' surface brightness profile of Abell 15 convolved 
to the 6$''$ resolution of the MIPS 24 \um\ image is plotted in 
dotted black curve. 
}
\end{figure}

\begin{figure}
\plotone{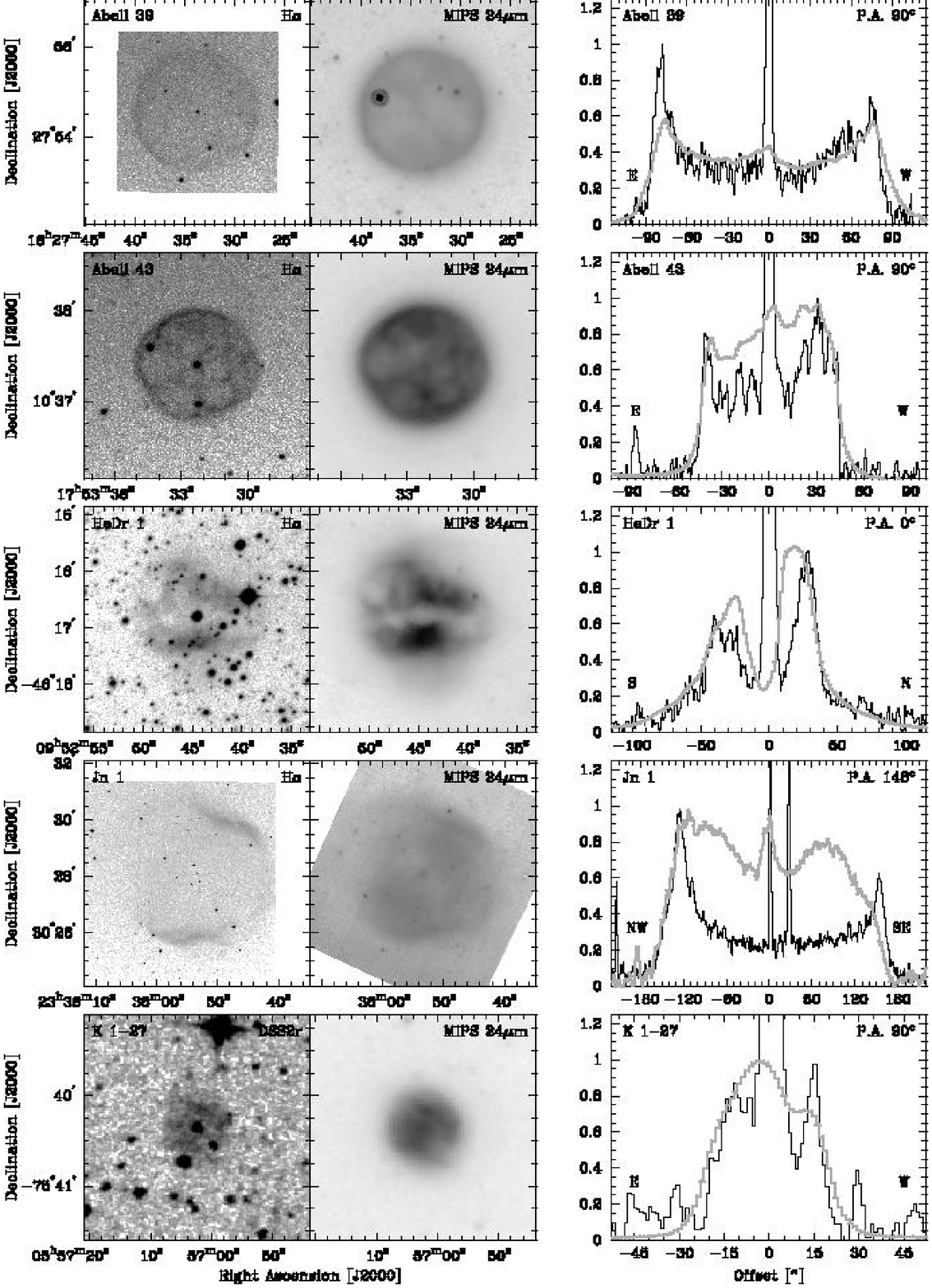}
\end{figure}
\begin{figure}
\plotone{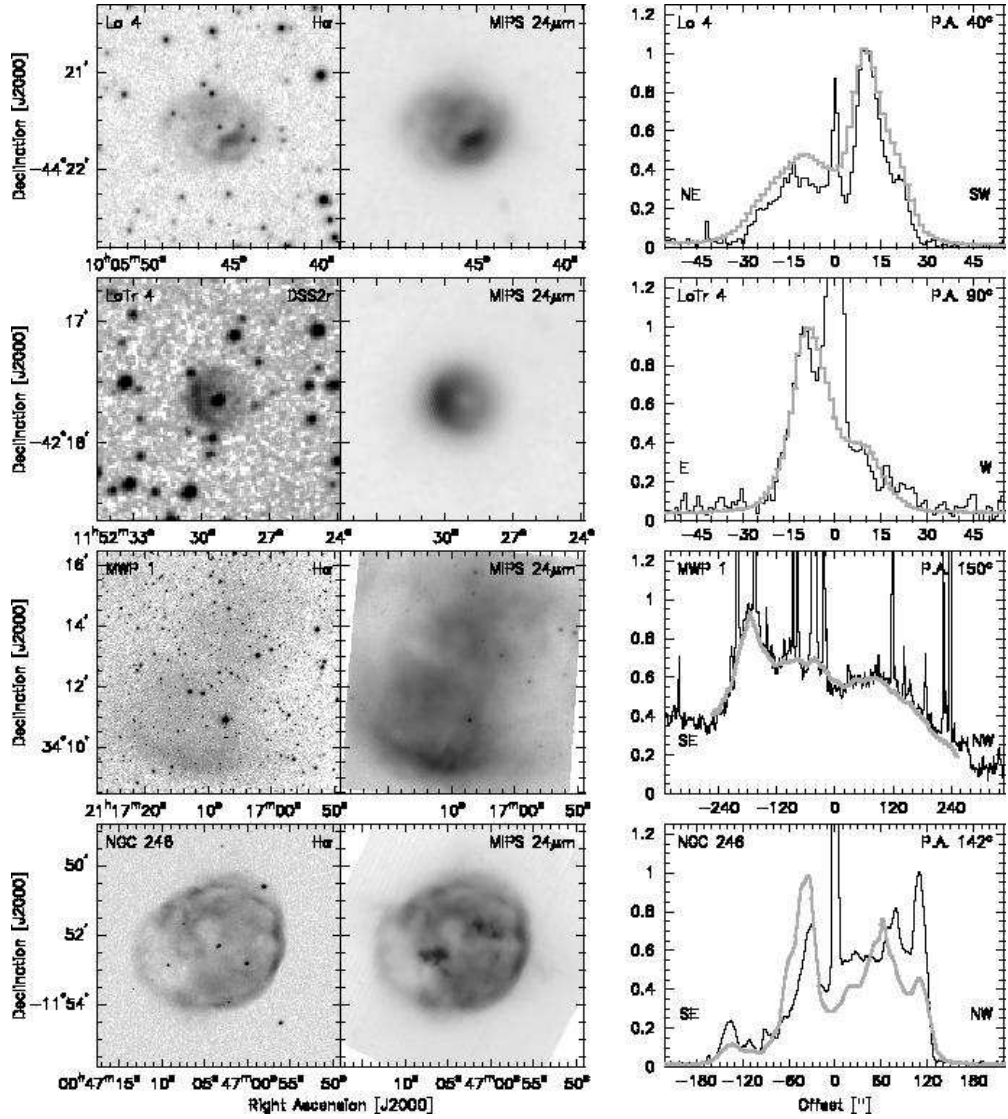}
\figurenum{2}
\label{fig2}
\caption{Same as Figure 1, but for Group 2 PNe.
 Note that the ``H$\alpha$'' surface brightness profiles of 
 K 1-27 and LoTr 4 are extracted from DSS2r images.
}
\end{figure}

\begin{figure}
\plotone{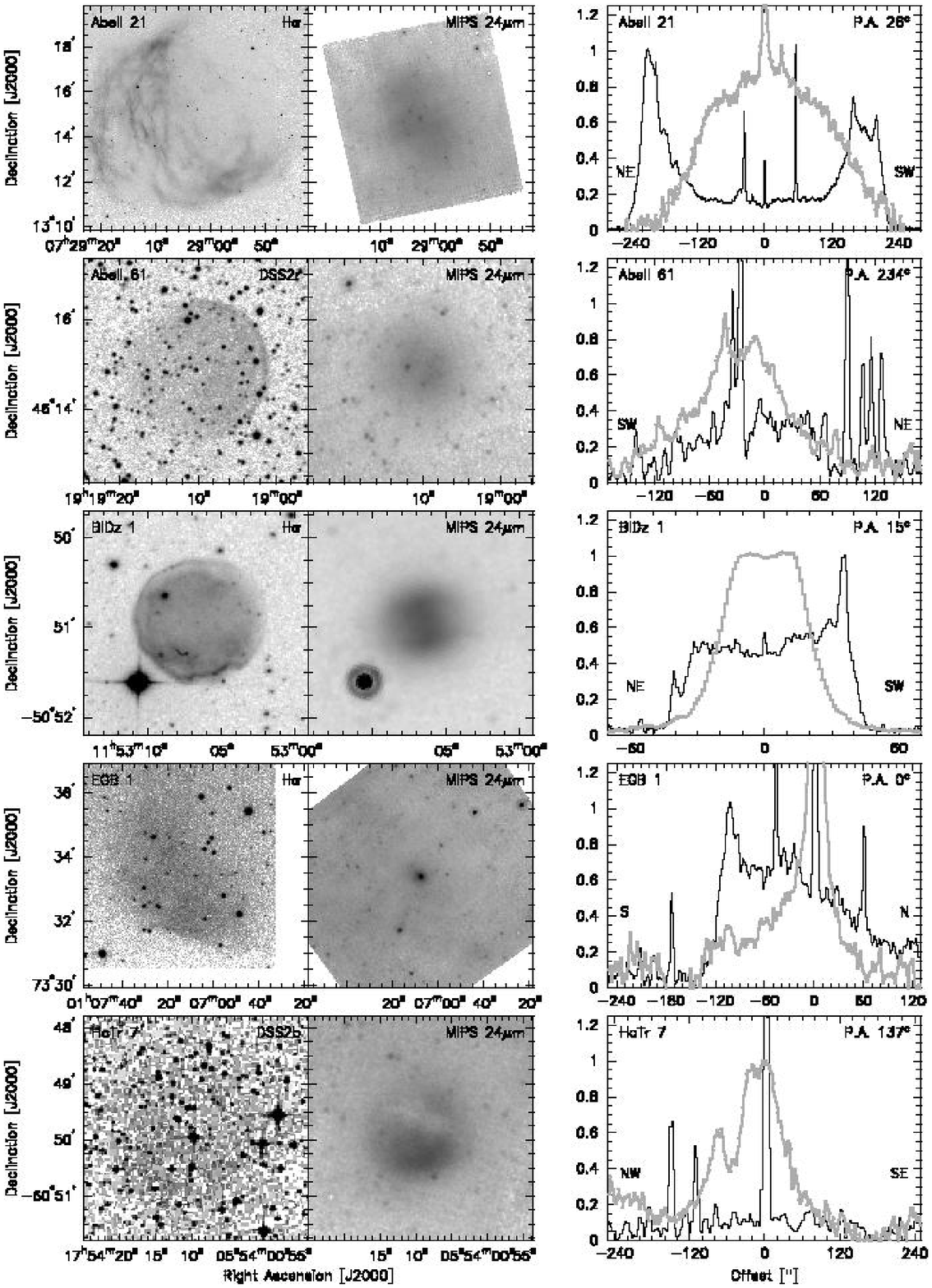}
\end{figure}
\begin{figure}
\plotone{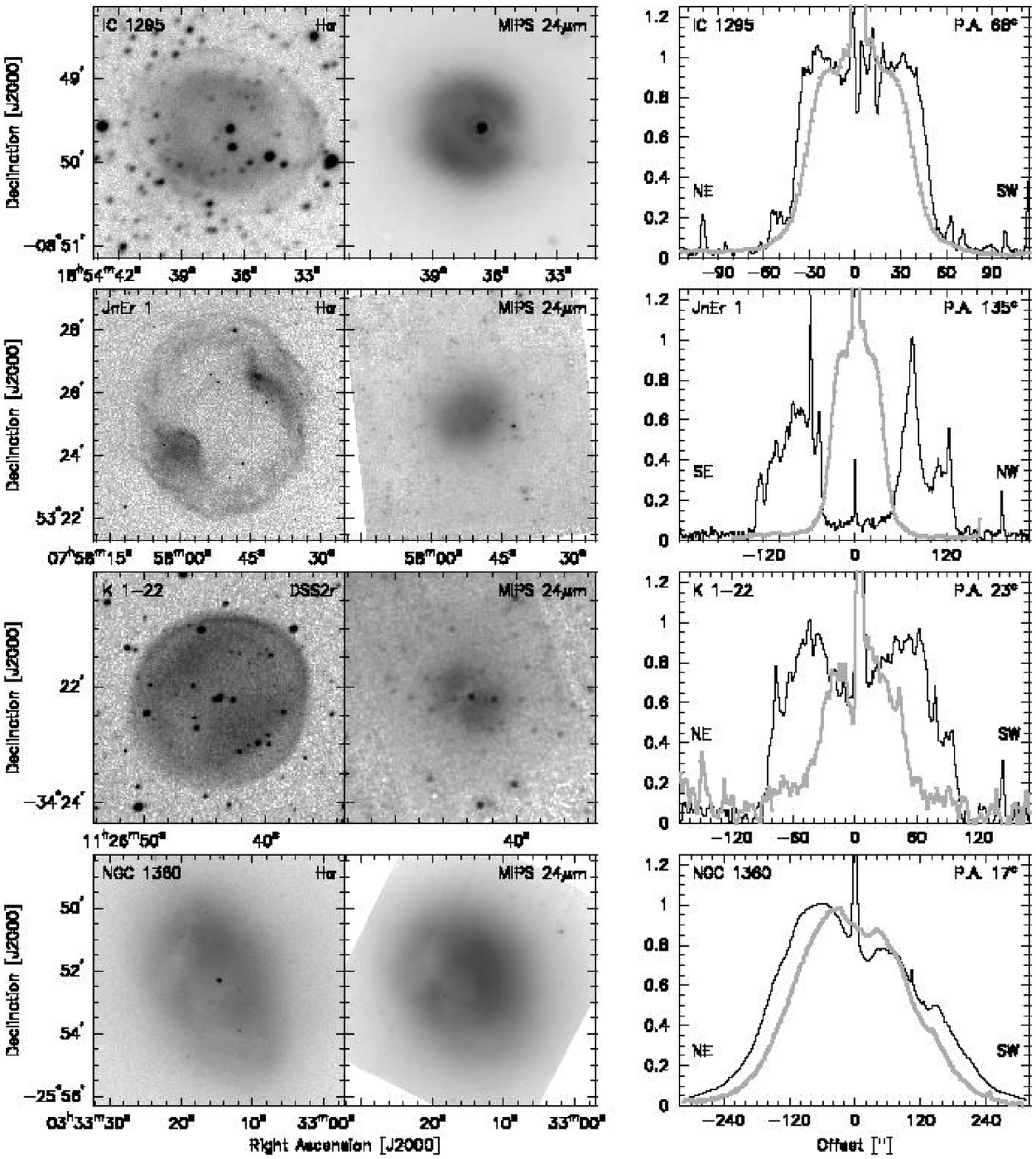}
\end{figure}
\begin{figure}
\plotone{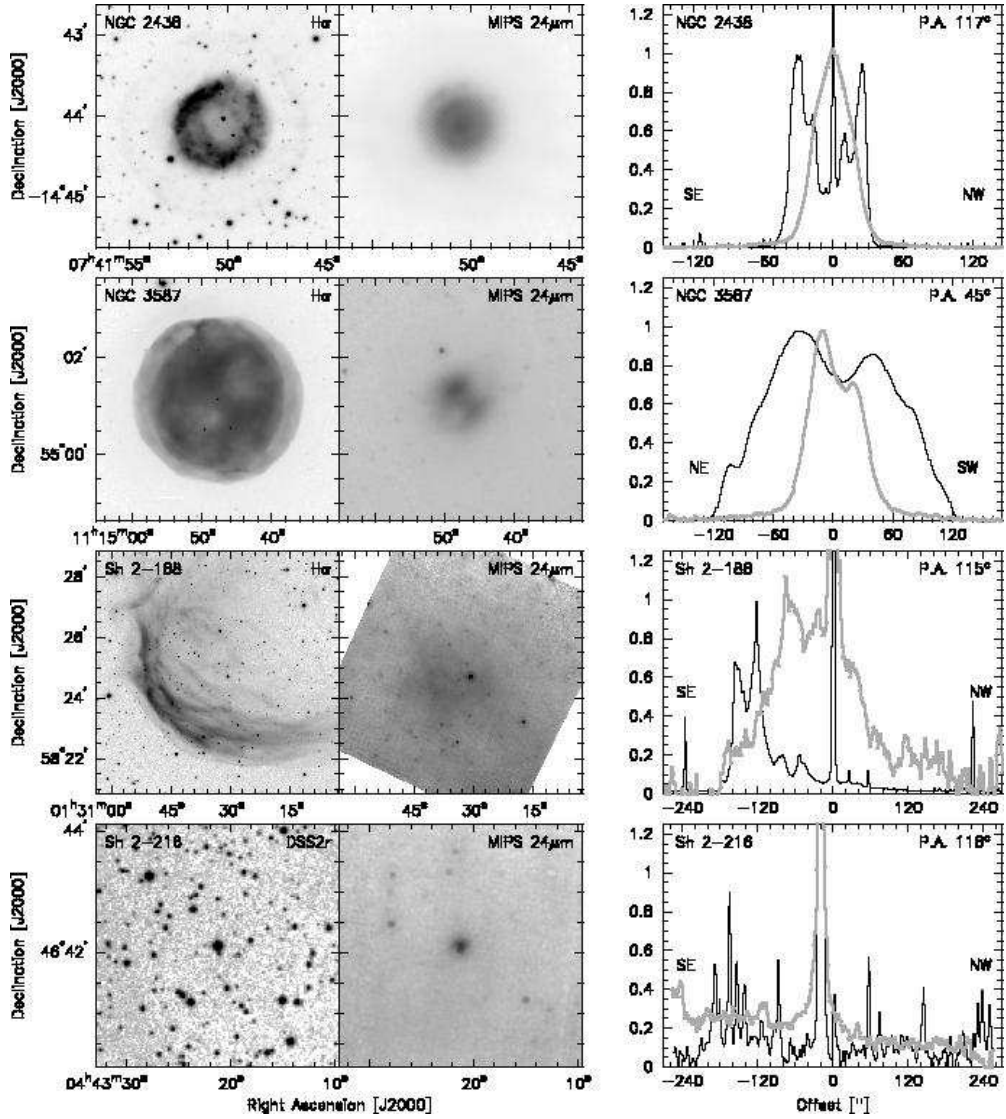}
\label{fig3}
\figurenum{3}
\caption{Same as Figure 1, but for Group 3 PNe.
Note that the ``H$\alpha$'' surface brightness profiles of 
Abell 61, K 1-22, and Sh 2-216 are extracted from DSS2r 
images, and HaTr 7 from a DSS2b image.
}
\end{figure}

\newpage
\begin{figure}
\figurenum{4}
\epsscale{1}
\plotone{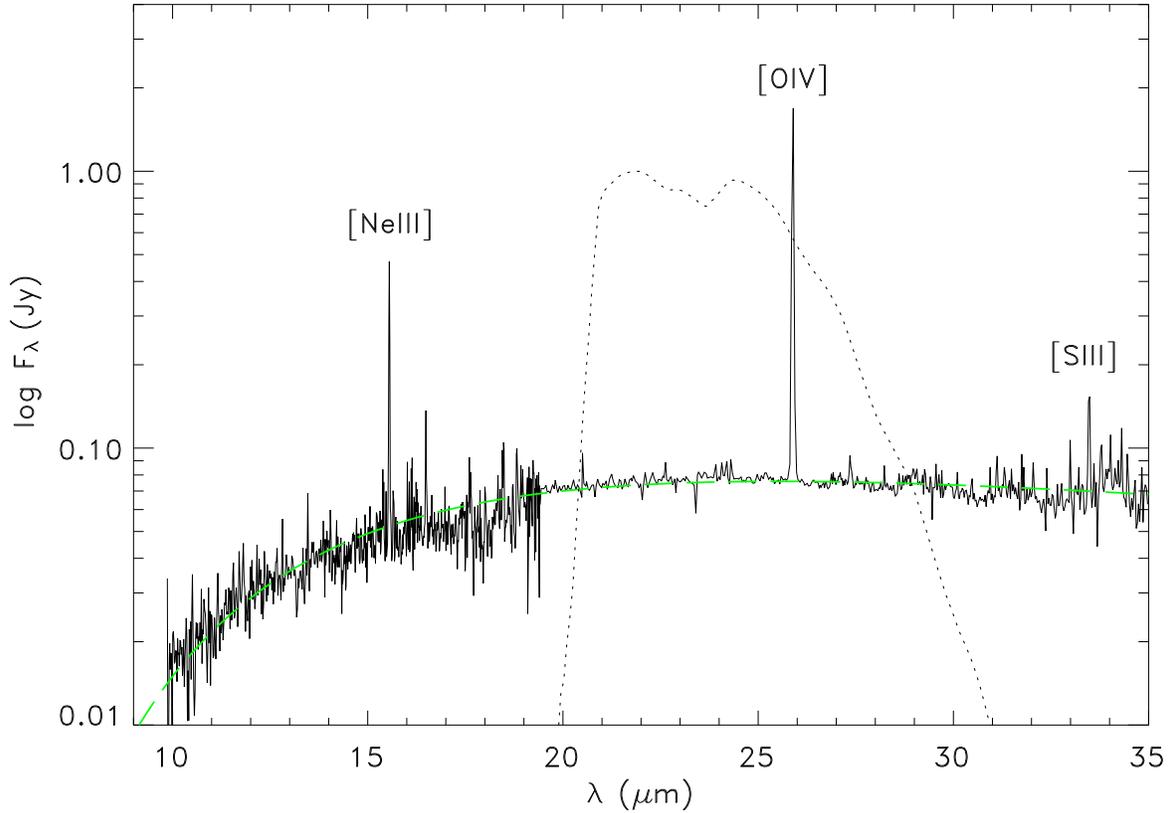}
\label{fig4}
\caption{ {\it Spitzer} IRS spectrum of the central
24 \um\ diffuse emission region of the Helix Nebula.
No off-source background spectrum was available for
zodiacal background subtraction; thus the continuum
in the spectrum is dominated by the zodiacal background,
which can be approximated by a 200 K blackbody model
(dashed curve).
The response curve of the MIPS 24 $\mu$m band is plotted
in a dotted curve.
}
\end{figure}

\newpage
\begin{figure}
\figurenum{5}
\epsscale{0.6}
\plotone{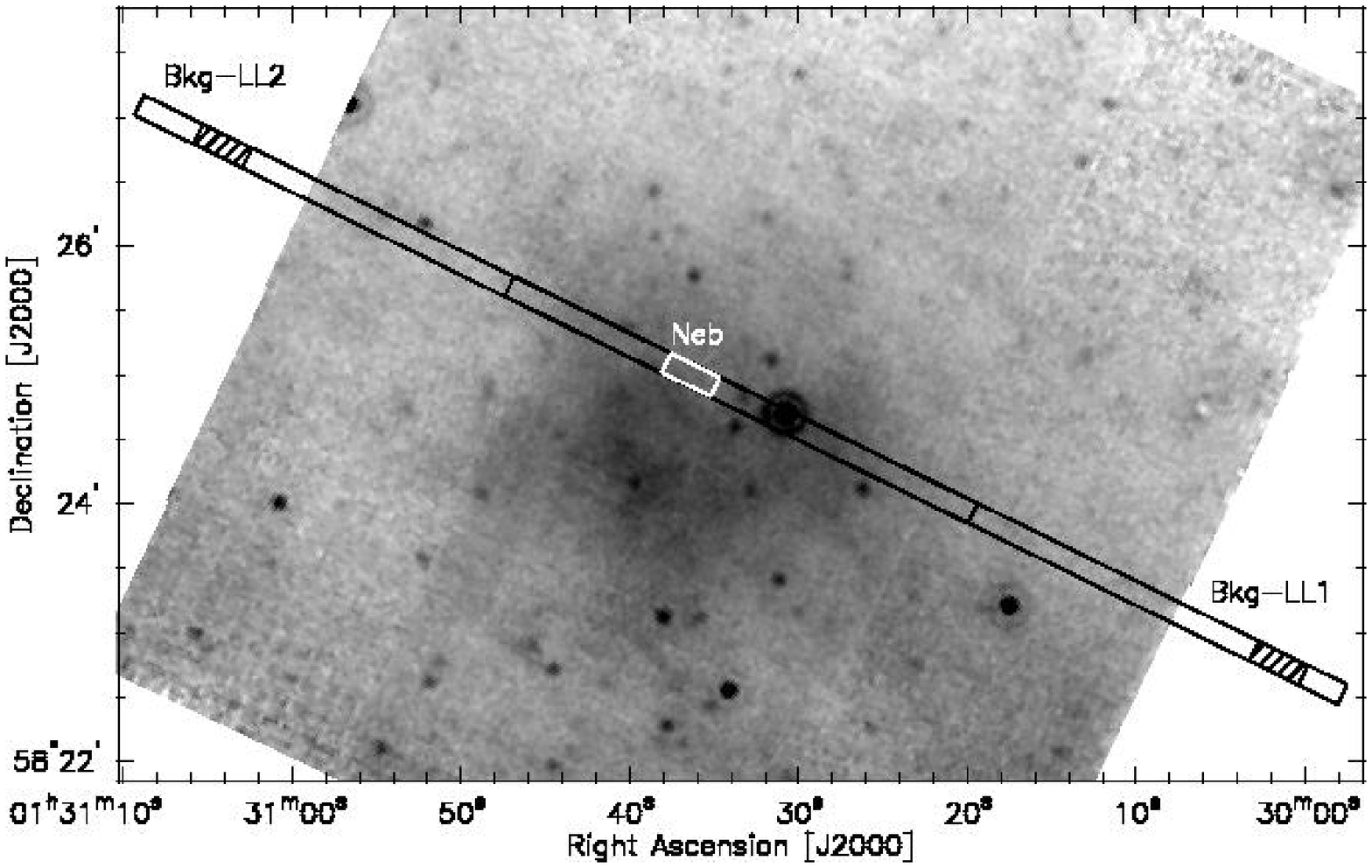}
\plotone{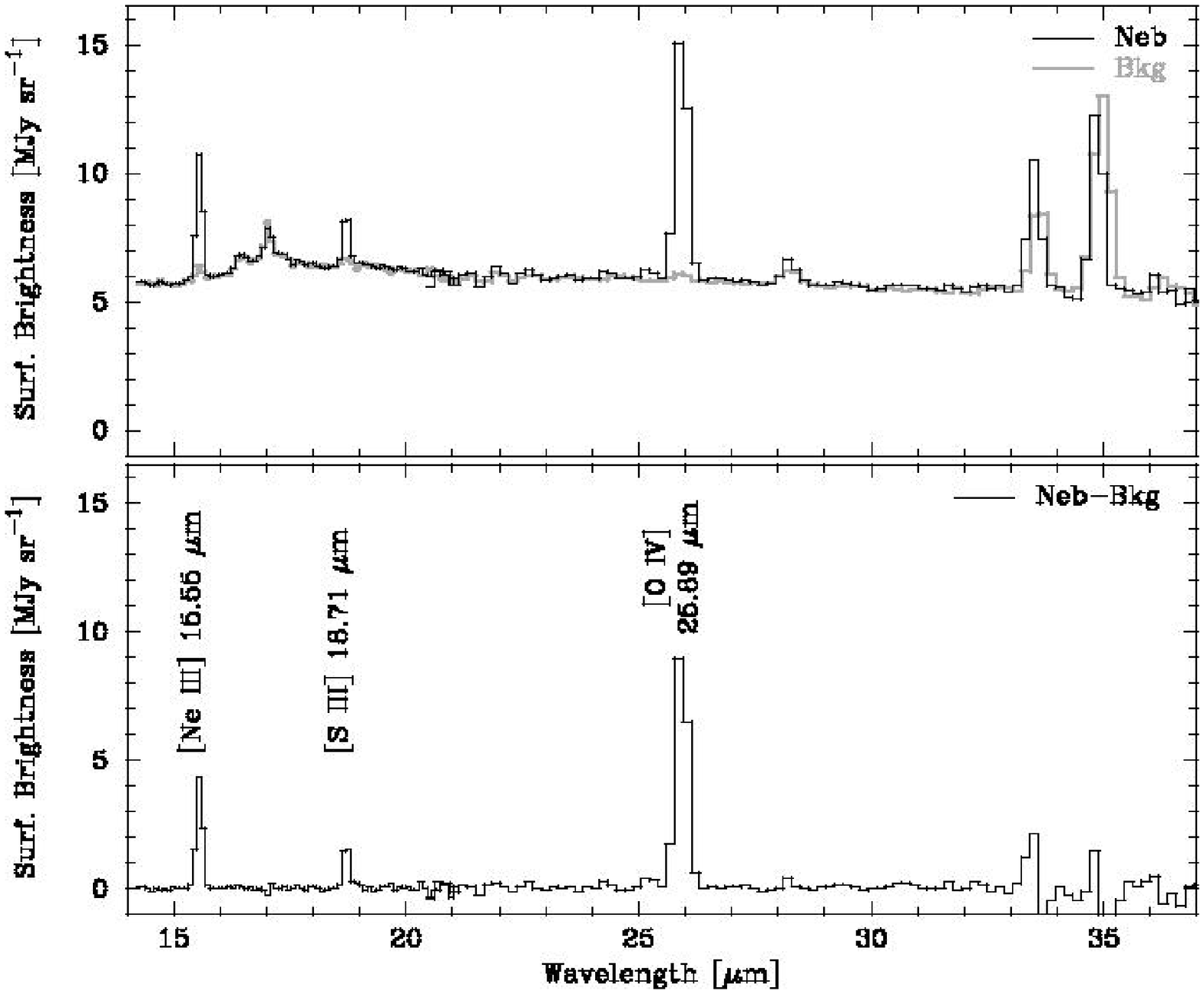}
\label{fig5}
\caption{{\it Spitzer} MIPS 24 \um\ image of Sh 2-188 (top panel)
 and IRS spectra  of the nebula (bottom panels).
 The nebular and background apertures are marked in the top panel.
 The LL1 and LL2 spectral modules have different spatial coverages,
 and thus the background apertures for LL1 and LL2 are different.
 In the upper spectral panel, the spectrum extracted from the nebular
 aperture ``Neb'' is plotted in black, and the background spectrum 
 (LL1 and LL2 combined) is plotted in grey.
 The background-subtracted nebular spectrum is plotted in the lower 
 spectral panel.  The CSPN Sh 2-188 is located near the center of the
 MIPS 24 \um\ image, project between a bright foreground star and 
 the nebular aperture.
 }
\end{figure}

\newpage
\begin{figure}
\figurenum{6}
\epsscale{0.5}
\begin{center}
    \epsfig{file=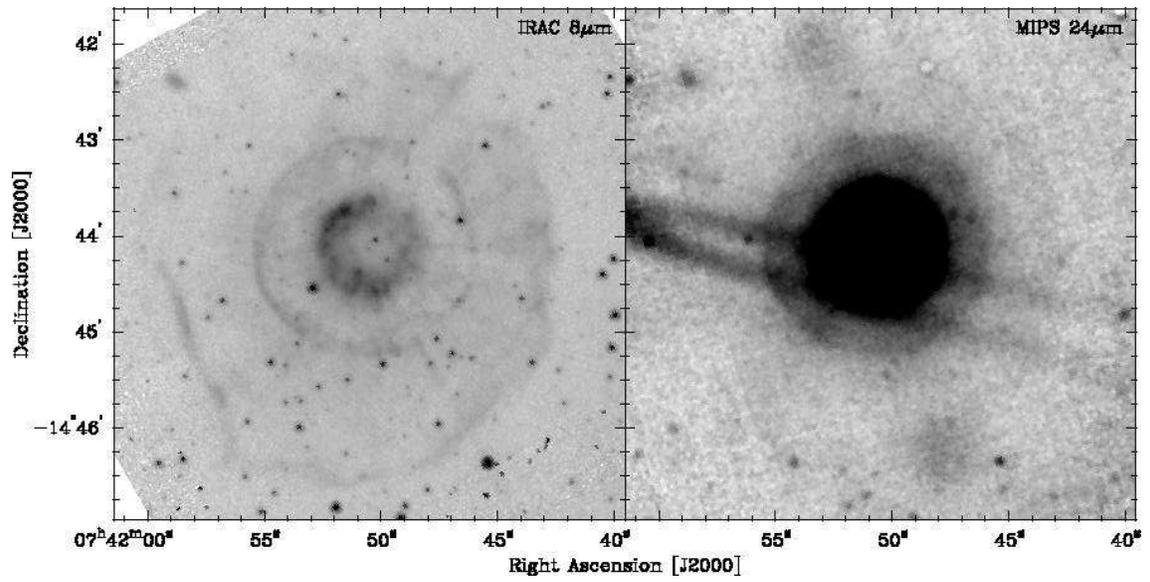, angle=-90, width=15cm}
  \end{center}
\label{fig6}
\caption{{\it Spitzer} IRAC 8.0 \um\ and MIPS 24 \um\
images of NGC 2438.  The two stripes of stray light
in the 24 \um\ image is spilled from the Calabash Nebula,
a very bright IR source in the post-AGB phase at $\sim$6.5$'$
northeast of NGC 2438.  The two patches of 24 \um\ emission 
at PAs = 20$^\circ$ and 200$^\circ$ outside the halo are caused 
by latent images of the bright main nebula.
}
\end{figure}

\newpage
\begin{figure}
\figurenum{7}
\begin{center}
    \epsfig{file=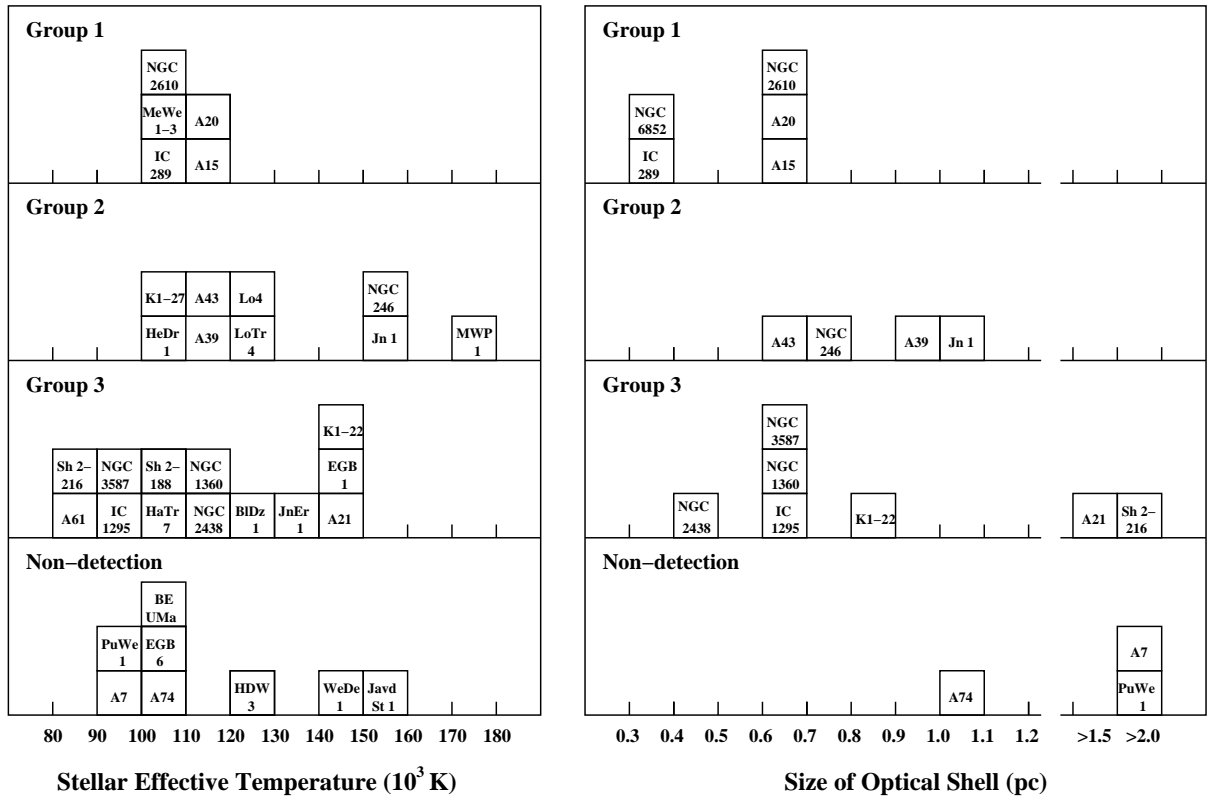, width=16 cm}
  \end{center}
\label{fig7}
\caption{Distributions of stellar effective temperatures of the
CSPNs and optical (H$\alpha$) shell sizes of the PNe.}
\end{figure}

\newpage
\begin{figure}
\figurenum{8}
\begin{center}
    \epsfig{file=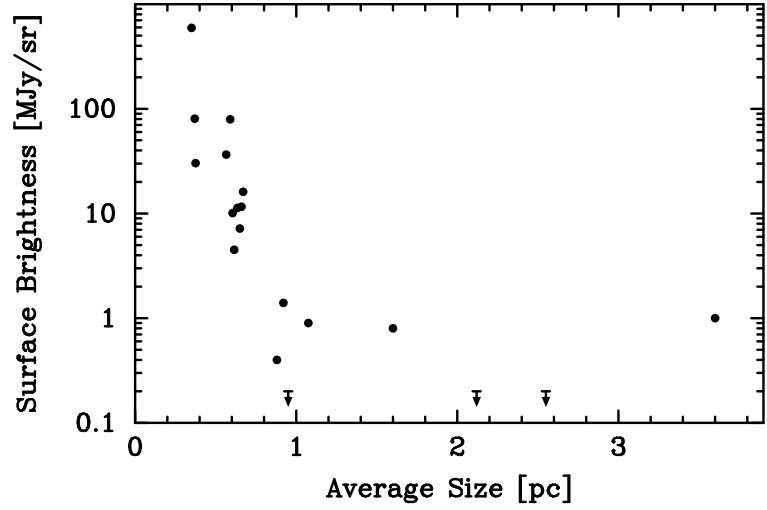, width=10cm}
  \end{center}
\label{fig8}
\caption{Plot of 24 \um\ surface brightness versus nebular size of PNe
that have distances available.  The average nebular size is measured 
from H$\alpha$ images (Column 7 of Table 1).  For non-spherical 
PNe, the average of the major and minor axes is used.}
\end{figure}

\begin{deluxetable}{llcccrclr}
\tabletypesize{\scriptsize}
\tablewidth{0pc}
\tablecaption{Survey Sample of Hot White Dwarfs and Pre-White Dwarfs}
\tablehead{
~~~PN~~         & ~~~PNG         & CSPN  & $T_{\rm eff}$\tablenotemark{a}  
& Size& Dist.\tablenotemark{b} &  Size& HeII/H$\beta$\tablenotemark{c} & 24 $\mu$m SB \\
~~Name     & ~Number      & Name & (10$^3$ K)   & (arcsec)  &  (pc)    & (pc) & flux ratio & 
(MJy/sr) \\
~~~(1)~~    & ~~~~(2)~~    & (3)  &   (4)        &   (5)     &  (6)~    & (7)  &  ~~~(8)  &  (9)~~~ 
}
\startdata
{\bf Group 1} &   &  &  &  &  &  & \\
Abell 15    & 233.5$-$16.3&  ...        & 110 & 36$\times$32  & 3656 & 0.64$\times$0.57& ~~~1.3 & 10.1~~~~\\
Abell 20    & 214.9+07.8  &  ...        & 119 & 68            & 1959 & 0.65  &  ~~~1.5 & 7.2~~~~\\
IC 289      & 138.8+02.8  &  ...        & 100 & 50            & 1434 & 0.35 &  ~~~0.88 & 593.4~~~~\\
MeWe 1-3    & 308.2+07.7  &  ...        & 100 & 18            & ...  & ... &  ~~~... & 27.9~~~~\\
NGC 2610    & 239.6+13.9  &  ...        & 100 & 59$\times$52  & 2194 & 0.63$\times$0.55& ~~~1.04 & 79.5~~~~\\
NGC 6852    & 042.5$-$14.5& WD 1958+015 & ... & 28            & 2710 & 0.37  & ~~~1.2 & 80.5~~~~\\
\hline
{\bf Group 2} &   &  &  &  &  &  & \\
Abell 39    & 047.0+42.4  & WD 1625+280 & 117 & 164           & 1163 & 0.92  & ~~~0.69 & 1.4~~~~\\
Abell 43    & 036.0+17.6  & WD 1751+106 & 117 & 85$\times$76  & 1619 & 0.67$\times$0.60&~~~0.93 & 11.3~~~~\\
HeDr 1      & 273.6+06.1  & LS 1362     & 100 & 90            & ...  & ...  & ~~~... & 13.4~~~~\\
Jn 1        & 104.2$-$29.6& WD 2333+301 & 150 & 330$\times$295&  709 & 1.13$\times$1.02& ~~~0.5 & 0.9~~~~\\
K 1-27      & 286.8$-$29.5& ...         & 105 & 44            & ...  & ... &~~~... & 8.7~~~~\\
Lo 4        & 274.3+09.1  & WD 1003$-$441&120 & 44            & ...  & ... & ~~~0.9 & 8.8~~~~\\
LoTr 4      & 291.4+19.2  & ...         & 120 & 30$\times$26  & ...  & ... & ~~~... & 17.9~~~~\\
MWP 1       & 080.3$-$10.4& WD 2115+339 & 170 & 300           & ...  & ... & ~~~... & 4.4~~~~\\
NGC 246	    & 118.8$-$74.7& WD 0044$-$121&150 & 273$\times$224&  470 & 0.62$\times$0.51&~~~1.2 &36.6~~~~\\
\hline
{\bf Group 3} &   &  &  &  &  &  & \\
Abell 21    & 205.1+14.2  & WD 0726+133 & 140 & 685$\times$530&  541 & 1.8$\times$1.4 & ~~~0.26 & 0.8~~~~\\
Abell 61    & 077.6+14.7  & WD 1917+461 & ~88  & 187           & ...  & ... & ~~~... & 0.6~~~~\\
BlDz 1      & 293.6+10.9  & ...         & 128 & 88$\times$85  & ...  & ... & ~~~...& 2.6~~~~\\
EGB 1       & 124.0+10.7  & WD 0103+732 & 147 & 300$\times$180& ...  & ... & ~~~...& 0.6:~~~\\
HaTr 7      & 332.5$-$16.9& ...         & 100 & 198$\times$180& ...  & ... & ~~~...& 0.9~~~~\\
IC 1295     & 025.4$-$04.7& WD 1851$-$088&~90 & 150$\times$121& 1024 & 0.74$\times$0.60& ~~~0.5& 16.1~~~~\\
JnEr 1      & 164.8+31.1  & WD 0753+535 & 130 & 425$\times$360& ...  & ... & ~~~0.26 & 1.1~~~~\\
K 1-22      & 283.6+25.3  & ...         & 141 & 184           &  988 & 0.88&~$<$0.1 & 0.4~~~~\\
NGC 1360    & 220.3$-$53.9& ...         & 110 & 460$\times$320&  348 & 0.78$\times$0.54&~~~1.0 &11.6~~~~\\
NGC 2438    & 231.8+04.1  & ...         & 114 & 69$\times$60  & 1203 & 0.4$\times$0.35&~~~0.4 & 30.3~~~~\\
NGC 3587    & 148.4+57.0  & WD 1111+552 & ~94 & 208$\times$198&  615 & 0.63$\times$0.60&~~~0.14 & 4.5~~~~\\
Sh 2-188    & 128.0$-$04.1& WD 0127+581 & 102 & 550           & ...  & ... & ~~~... & 0.6~~~~\\
Sh 2-216    & 158.5+00.7  & WD 0439+466 & ~83 & 5760          &  129 & 3.6 & ~~~... & 1.0:~~~\\
\hline
\multicolumn{2}{l}{\bf Not Detected}   &  &  &  &  &  & \\
Abell 7     & 215.5$-$30.8& WD 0500$-$156&~99 & 870$\times$672&  676 & 2.9$\times$2.2 & ~~~...& $<$0.2~~~~\\
Abell 74    & 072.7$-$17.1& WD 2114+239 & 108 & 870$\times$792&  752 & 1.0$\times$0.9 & ~~~...& $<$0.2~~~~\\
EGB 6       & 221.5+46.3  & WD 0950+139 & 100 & 834           & ...  & ... & ~~~... & $<$0.2~~~~\\
BE UMa      & 144.8+65.8  & BE UMa      & 105 & ...           & ...  & ... & ~~~... & $<$0.2~~~~\\
HDW 3       & 149.4$-$09.2& WD 0322+452 & 125 & 540           & ...  & ... & ~~~... & $<$0.2~~~~\\
JavdSt 1    & 085.4+52.3  & WD 1520+525 & 150 & 660           & ...  & ... & ~~~... & $<$0.2~~~~\\
PuWe 1      & 158.9+17.8  & WD 0615+556 & ~94 & 1200          &  365 & 2.12& ~~~... & $<$0.2~~~~\\
WeDe 1      & 197.4$-$06.4& WD 0556+106 & 141 &1020$\times$840& ...  & ... & ~~~... & $<$0.2~~~
\enddata
\tablenotetext{a}{The stellar effective temperatures are either derived or compiled
  by \citet{Nap99}, except those of Abell 21, Jn 1, and JnEr 1, which are from
  \citet{WH06}.}
\tablenotetext{b}{The distances are from \citet{Cahn92}, except those of Abell 7, Abell 21, 
Abell 74, PuWe 1, and Sh 2-216 that are from \citet{Harris07}.}
\tablenotetext{c}{The \ion{He}{2} $\lambda$4686/H$\beta$ flux ratios are from \citet{Tylenda94}.}
\end{deluxetable}


\begin{thebibliography}{}

\bibitem[Beintema et al.(1996)]{Betal96} 
Beintema, D.~A., et al.\ 1996, \aap, 315, L253

\bibitem[Bernard Salas et al.(2001)]{BS01} 
Bernard Salas, J., Pottasch, S.~R., Beintema, D.~A., \& Wesselius, 
P.~R.\ 2001, \aap, 367, 949

\bibitem[Cahn et al.(1992)]{Cahn92} 
Cahn, J.~H., Kaler, J.~B., \& Stanghellini, L.\ 1992, \aaps, 94, 399 

\bibitem[Corradi et al.(2000)]{Corradi00} 
Corradi, R.~L.~M., Sch{\"o}nberner, D., Steffen, M., \& 
Perinotto, M.\ 2000, \aap, 354, 1071 

\bibitem[Corradi et al.(2003)]{Corradi03} 
Corradi, R.~L.~M., Sch{\"o}nberner, D., Steffen, M., \& 
Perinotto, M.\ 2003, \mnras, 340, 417 

\bibitem[Dgani \& Soker(1998)]{DS98} 
Dgani, R., \& Soker, N.\ 1998, \apj, 495, 337 

\bibitem[Fazio et al.(2004)]{Fazio04}
Fazio, G.~G., et al.\ 2004, \apjs, 154, 10

\bibitem[Hajian et al.(1997)]{Hetal97}
Hajian, A.\ R., Frank, A., Balick, B., Terzian, Y.\ 1997, \apj,
477, 226

\bibitem[Harris et al.(2007)]{Harris07} 
Harris, H.~C., et al.\ 2007, \aj, 133, 631

\bibitem[Hartl \& Tritton(1985)]{HT85} 
Hartl, H., \& Tritton, S.~B.\ 1985, \aap, 145, 41

\bibitem[Hora(2009)]{Hora09}
Hora, J.~L.\ 2009, in Asymmetrical Planetary Nebulae IV,
eds.\ R.~L.~M.\ Corradi, A.\ Manchado, \& N.\ Soker,
Springer-Verlag, 27

\bibitem[Hora et al.(2004)]{Hora04} 
Hora, J.~L., Latter, W.~B., Allen, L.~E., Marengo, M., 
Deutsch, L.~K., \& Pipher, J.~L.\ 2004, \apjs, 154, 296

\bibitem[Hora et al.(2009)]{Hetal08} 
Hora, J.~L., Marengo, M., Smith, H.~A., Cerrigone, L., 
\& Latter, W.~B.\ 2009, in ``The Evolving ISM in the Milky 
Way and Nearby Galaxies,''  ArXiv e-prints, 803, 
arXiv:0803.3937

\bibitem[Jacoby \& van de Steene(1995)]{Jacoby95} 
Jacoby, G.~H., \& van de Steene, G.\ 1995, \aj, 110, 1285 

\bibitem[Kiss et al.(2008)]{Ketal08} 
Kiss, L.~L., Szab{\'o}, G.~M., Balog, Z., Parker, Q.~A., \& Frew, 
D.~J.\ 2008, \mnras, 391, 399 

\bibitem[Kwok(1983)]{Kwok83}
Kwok, S.\ 1983, in IAU Symp.\ 103, Planetary Nebulae, 
ed.\ D.~R.\ Flower (Dordrecht: Kluwer), 293

\bibitem[Liebert et al.(1995)]{Liebert95} 
Liebert, J., Tweedy, R.~W., Napiwotzki, R., \& Fulbright, 
M.~S.\ 1995, \apj, 441, 424 

\bibitem[Manchado et al.(1996)]{Man96}
Manchado, A., Guerrero, M.~A., Stanghellini, L., \& Serra-Ricart, 
M.\ 1996, The IAC Morphological Catalog of Northern Galactic
Planetary Nebulae (La Laguna, Spain: IAC)

\bibitem[McCook \& Sion(1999)]{MS99} 
McCook, G.~P., \& Sion, E.~M.\ 1999, \apjs, 121, 1

\bibitem[M{\'e}ndez(1978)]{Men78} 
M{\'e}ndez, R.~H.\ 1978, \mnras, 185, 647 

\bibitem[Napiwotzki(1999)]{Nap99} 
Napiwotzki, R.\ 1999, \aap, 350, 101

\bibitem[Parker et al.(2005)]{Petal05} 
Parker, Q.~A., et al.\ 2005, \mnras, 362, 689

\bibitem[Regan \& Gruendl(1995)]{RG95}
Regan, M. W. \& Gruendl, R. A.\ 1995, in ASP Conf.\ Ser.\ 77,
ADASS IV Proc., eds.\ R.\ A. Shaw, H.\ E.\ Payne, \& 
J. J. E. Hayes (San Francisco, CA: ASP), 335

\bibitem[Rieke et al.(2004)]{Rieke04}
Rieke, G.~H., et al.\ 2004, \apjs, 154, 25

\bibitem[Sch\"onberner(1983)]{Schon83} 
Sch\"onberner, D.\ 1983, \apj, 272, 708

\bibitem[Soker \& Livio(1989)]{SL89} 
Soker, N., \& Livio, M.\ 1989, \apj, 339, 268

\bibitem[Stanghellini et al.(2007)]{Stan07} 
Stanghellini, L., et al.\ 2007, \apj, 671, 1669

\bibitem[Su et al.(2004)]{Su04} 
Su, K.~Y.~L., et al.\ 2004, \apjs, 154, 302

\bibitem[Su et al.(2007)]{Su07} 
Su, K.~Y.~L., et al.\ 2007, \apjl, 657, L41 

\bibitem[Tylenda et al.(1994)]{Tylenda94} 
Tylenda, R., Stasi{\'n}ska, G., Acker, A., \& Stenholm, B.\ 1994, 
\aaps, 106, 559 

\bibitem[Ueta(2006)]{Ueta06} 
Ueta, T.\ 2006, \apj, 650, 228

\bibitem[Villaver et al.(2003)]{Vetal03} 
Villaver, E., Garc{\'{\i}}a-Segura, G., \& Manchado, A.\ 2003, 
\apjl, 585, L49 

\bibitem[Volk et al.(2007)]{Vetal07} 
Volk, K., Kwok, S., \& Hrivnak, B.~J.\ 2007, \apj, 670, 1137

\bibitem[Werner \& Herwig(2006)]{WH06} 
Werner, K., \& Herwig, F.\ 2006, \pasp, 118, 183

\bibitem[Werner et al.(2004)]{Werner04}
Werner, M.~W., et al.\ 2004, \apjs, 154, 1

\end{thebibliography}
\end{document}